\documentclass[12pt]{iopart}
\usepackage{graphicx}   	
\usepackage{verbatim}  	 	
\usepackage{color}      	
\usepackage{hyperref}   	
\hypersetup{pdfborder = {0 0 0},colorlinks=true}
\newcommand{\sh}{\mathrm{sh}}
\newcommand{\ch}{\mathrm{ch}}
\renewcommand{\th}{\mathrm{th}}
\newcommand{\arcsh}{\mathrm{arcsh}}
\newcommand{\arcth}{\mathrm{arcth}}

\begin{document}
\title{Bethe-Peierls approximation and the inverse Ising model}


\author{H. Chau Nguyen and Johannes Berg}
\address{University of Cologne, Institute for Theoretical Physics\\Z\"{u}lpicher Stra{\ss}e 77, 50937 K\"{o}ln, Germany}
\eads{\mailto{cnguyen@thp.uni-koeln.de} and \mailto{berg@thp.uni-koeln.de}}



\begin{abstract}
We apply the Bethe-Peierls approximation to the problem of the inverse Ising model and show how the linear response relation leads to a simple method to reconstruct couplings and fields of the Ising model.  This reconstruction is exact on tree graphs, yet its computational expense is comparable to other mean-field methods. We compare the performance of this method to the independent-pair, naive mean-field, Thouless-Anderson-Palmer approximations, the Sessak-Monasson expansion, and susceptibility propagation in the Cayley tree, SK-model and random graph with fixed connectivity. At low temperatures, Bethe reconstruction outperforms all these methods, while at high temperatures it is comparable to the best method available so far (Sessak-Monasson). The relationship between Bethe reconstruction and other mean-field methods is discussed.
\end{abstract}



\maketitle

The enormous and ongoing growth of data in molecular biology has led to a surge of interest in inverse problems. Examples are the reconstruction of gene regulatory networks from expression levels~\cite{Baily}, the identification of neural interactions from neural spike recordings~\cite{Cocco_PNAS}, and the  prediction of protein-protein interactions from the evolutionary correlation between amino acids~\cite{Weigt}. The paradigm of such inverse problems is the inverse Ising model, where the parameters of the Ising model Hamiltonian (couplings and fields) are to be inferred from a set of spin configurations drawn independently from the equilibrium measure. The goal is to use correlations observed between spins to infer the underlying, unknown couplings and fields. This problem is intrinsically hard, as pairs of spins can exhibit correlations without interacting directly with each other.

In this note, we consider the inverse Ising model at the level of the Bethe-Peierls approximation (BP) and show how the linear response approach~\cite{Kappen,Welling_AI,Opper} leads to a reconstruction of the Ising model that is efficient, straightforward and outperforms currently available mean-field-like methods in benchmarks for strong couplings (and does as well as they do at weak couplings).

Consider the Boltzmann measure
\begin{equation}
p[\mathbf{s}] = \frac{1}{Z} \mathrm{e} ^{-H[\mathbf{s}]} \label{eq: Boltzmann}
\end{equation}
of the Ising model with $H [\mathbf{s}] = - \sum_{i} h_i s_i - \sum_{(ij)} J_{ij} s_i s_j$, specifying the statistical weight of a configuration $\mathbf{s} = (s_1, s_2,..., s_N)$ given couplings $J_{ij}$ and local fields $h_i$  (at inverse temperature $\beta = 1$), where $Z$ is the partition function. The indices $(ij)$ run over all pairwise interactions. The inverse Ising problem is then to recover the couplings and fields from a given set of $M$ spin configurations $D = \{ \mathbf{s}^1, \mathbf{s}^2,..., \mathbf{s}^{M} \}$ drawn independently from (\ref{eq: Boltzmann}). The log-likelihood of couplings and fields given such a set of observations, normalized by $M$, is
\begin{equation}
L(\{ s_{i}^{\mu} \}|\{h_i, J_{ij}\}) = - \ln Z [\{h_i, J_{ij}\}] + \sum_{i} {h_i} \frac{1}{M}\sum_{\mu} s_{i}^{\mu} +  \sum_{(ij)} J_{ij} \frac{1}{M}\sum_{\mu} s_{i}^{\mu} s_{j}^{\mu} \ .
\label{eq: likelihood}
\end{equation}
Maximizing the log-likelihood with respect to the Ising model parameters $J_{ij}$ and $h_i$ leads to
\begin{eqnarray}
m_i (\{h_i,J_{ij}\}) & = & m_i^D \ , \nonumber
\\
C_{ij} (\{h_i,J_{ij}\}) & = & C_{ij}^D \label{eq: match} \ ,
\end{eqnarray}
where $m_i = \langle s_i \rangle$ and $C_{ij} = \langle s_i s_j \rangle - \langle s_i \rangle  \langle s_j \rangle $ are the magnetizations and connected correlations under the Boltzmann distribution (\ref{eq: Boltzmann}). The right-hand sides are the corresponding quantities estimated from data,  $m_i^{D} = \frac{1}{M}\sum_{\mu} s_{i}^{\mu} $ and $C_{ij}^{D} = \frac{1}{M}\sum_{\mu} s_{i}^{\mu} s_{j}^{\mu} - m_i^D m_j^D$ \cite{MacKay}.

Many different approaches to find the couplings $J_{ij}$ and fields $h_i$ by maximizing the log-likelihood (\ref{eq: likelihood}) or by directly solving the self-consistent equations (\ref{eq: match}) have been taken, including gradient descent with Monte-Carlo simulation~\cite{MacKay}, independent-pair approximation (IP)~\cite{Roudi_FCN}, naive mean-field (MF)~\cite{Kappen}, Thouless-Anderson-Palmer approximation (TAP)~\cite{Kappen,TAP}, Sessak-Monasson  expansion (SM)~\cite{Sessak}, susceptibility propagation (SusP)~\cite{Welling_NC,Mezard_JOP} and more recently an adaptive cluster expansion~\cite{Cocco_PRL} and pseudo-likelihood maximization~\cite{Ravikumar,Aurell}. The canonical way to use a mean-field approximation for the inverse problem is to calculate the correlations by linear response  of the magnetizations $m_i$ to changes of the local fields $h_i$ to yield the connected correlations and solve for the couplings. In the same way, our approach is based on the Bethe-Peierls approximation~\cite{Bethe} combined with the linear response relation.

We begin by deriving the connected correlation for a system of spins in a tree graph, where the Boltzmann distribution (\ref{eq: Boltzmann}) can be written exactly in terms of the Bethe ansatz, that is the combination of the one-point marginals $b_i$ at the spins and the two-point marginals $b_{ij}$ at the bonds,
\begin{equation}
p_{BP} \left[ \mathbf{s} \right] = {\prod_{i} b_i(s_i)^{1-z_i}}{\prod_{(i,j)} b_{ij}(s_i, s_j)} \ , \label{eq: bp}
\end{equation}
where $z_i$ is the number of non-zero bonds connected to spin $i$ \cite{Bethe,Kikuchi,Mezard,Yedidia_22,Mean_Field}. The marginals $b_i$ and  $b_{ij}$, which are still unknown, can be parameterized by local magnetizations $m_i$ and correlation parameters $\bar{C}_{ij}$ as
\begin{eqnarray}
b_i(s_i) =\frac{1}{2} (1 + m_i s_i) \ , \nonumber
\\
b_{ij} (s_i, s_j) = \frac{1}{4} [(1 + m_i s_i)(1 + m_j s_j) + \bar{C}_{ij} s_i s_j] \ , \label{eq: bp para}
\end{eqnarray}
with constraints
\begin{eqnarray}
{-1} \leq m_i \leq {+1} \ , \nonumber
\\
{-1}+\vert m_i + m_j \vert -m_im_j  \leq  \bar{C}_{ij}  \leq  +1 - \vert m_i - m_j \vert - m_i m_j \ . \label{eq: constraint}
\end{eqnarray}

Instead of \textit{single spins} coupled to effective fields in naive mean-field theory, the Bethe ansatz is based on \textit{bonds}.  It is thus a natural approximation to take in the context of the inverse problem, where it is bonds that are to be determined, not the statistics of spins.

The self-consistent equations for the parameters $m_i$ and $\bar{C}_{ij}$ describing the statistics of spin pairs in equation (\ref{eq: bp para}) can be found by minimizing the Kullback-Leibler divergence $\mathcal{D}$ between the Bethe ansatz (\ref{eq: bp})  and the Boltzmann measure (\ref{eq: Boltzmann}),  $\mathcal{D}(p_{BP}[\mathbf{s}],p[\mathbf{s}]) = \sum_{\mathbf{s}} p_{BP}[\mathbf{s}] \ln ( {p_{BP}[\mathbf{s}]}/{p[\mathbf{s}]} )$, giving
\begin{eqnarray}
\fl
h_i + \sum_{j \in \partial i} J_{ij} m_j &=& (1-z_i) \arcth{(m_i)} + \nonumber \\
&& +\sum_{j \in \partial i}  \sum_{s_i, s_j} \frac{s_i + m_j s_i s_j}{4} \ln \frac{(1+m_i s_i)(1+m_j s_j) + \bar{C}_{ij} s_i s_j}{4}\ , \label{eq: bethe m}
\\
J_{ij} &=&\sum_{s_i, s_j} \frac{s_i s_j}{4} \ln \frac{(1+m_i s_i)(1+m_j s_j)+\bar{C}_{ij} s_i s_j}{4} \ , \label{eq: bethe chi}
\end{eqnarray}
where $\partial i$ stands for the boundary set containing all spins that are connected to spin $i$ by a bond.

The self-consistent equation (\ref{eq: bethe chi}) is quadratic in $\bar{C}_{ij}$, but only one of its solutions is compatible with the constraint (\ref{eq: constraint}), which is
\begin{equation}
\bar{C}_{ij} = \frac{(1- m_i^2 - m_j^2) t_{ij} + 2 m_i m_j}{1+ \sqrt{D_{ij}}} - m_i m_j \ , \label{eq: bethe chi 1}
\end{equation}
where $D_{ij} = 1 - 2 m_i m_j t_{ij} -(1- m_i^2 - m_j^2) t_{ij}^2$ and  $t_{ij}=\th (2 J_{ij})\ .$

The correlation parameters $\bar{C}_{ij}$ give the connected correlation for pairs of spins that interact with each other in the tree. To find the connected correlation $C_{ij}$ between \emph{any pair} of spins, we follow a route based on the linear response relation~\cite{Kappen,Welling_AI,Mean_Field}. Differentiating the self-consistent equations (\ref{eq: bethe m}) and  (\ref{eq: bethe chi 1}) with respect to the fields $h_i$ yields a set of equations for the susceptibilities ${\partial m_{i}}/{\partial h_{j}}$. The general linear response relation $C_{ij} = {\partial m_{i}}/{\partial h_{j}}$ then directly gives the connected correlations, for which we obtain
\begin{eqnarray}
(C^{-1})_{ij} &=& -J_{ij} + \tilde{J}_{ij} + \frac{\bar{C}_{ij}}{(\bar{C}_{ij})^2-(1-m_i^2)(1-m_j^2)} \ ,\  (i \ne j) \ ,
\label{eq: bethe Cij}
\\
(C^{-1})_{ii} &=& \frac{1-z_i}{1-m_i^2} - \sum_{j \in \partial i} \frac{1-m_j^2}{(\bar{C}_{ij})^2-(1-m_i^2)(1-m_j^2)} \ ,
\label{eq: bethe Cii}
\end{eqnarray}
with
\begin{equation}
\tilde{J}_{ij} = \frac{1}{4} \mathrm{ln} \left\{ \frac{[ (1+m_i)(1+m_j) + \bar{C}_{ij}] [ (1-m_i)(1-m_j) + \bar{C}_{ij}]}{[(1+m_i)(1-m_j) - \bar{C}_{ij}] [ (1-m_i)(1+m_j)- \bar{C}_{ij}]} \right\} \ ,
\label{eq: IP}
\end{equation}
where the coupling matrix $J_{ij}$ was extended to every pair of spins so that $J_{ij} = 0$ if $i$ and $j$ are not connected by a bond.

Note that the first two terms in the right hand side of equation (\ref{eq: bethe Cij}) cancel exactly due to (\ref{eq: bethe chi}), we however keep them to relate the expression to the Sessak-Monasson expansion later on.

An equivalent version of equations (\ref{eq: bethe Cij})-(\ref{eq: IP}) was derived by Welling and Teh in the context of the forward problem~\cite{Welling_AI}, and was used to estimate the data evidence in Bayesian inference~\cite{Welling_NIPS}. These equations can be used to estimate the couplings and local fields in the inverse Ising problem. To do so, we first assume a fully connected model without self-interactions. It follows that $z_i = N-1$, where $N$ is the number of spins. The Bethe ansatz, together with the equations (\ref{eq: bethe m}), (\ref{eq: bethe chi}), (\ref{eq: bethe Cij}) and (\ref{eq: bethe Cii}), is exact only in tree graphs, using it as an approximation in loopy graphs is usually referred to as Bethe-Peierls approximation. In loopy graphs, the correlation between interacting spins estimated by the linear response relation (\ref{eq: bethe Cij}) is different from  the correlation parameter (\ref{eq: bethe chi 1}), which is, in general, of lower accuracy \cite{Opper}. In equation (\ref{eq: bethe Cij}), we identify the magnetizations $m_i$ and the connected correlations $C_{ij}$  with the values $m_i^D$ and $C_{ij}^D $  estimated from data. Inserting the solution  (\ref{eq: bethe chi 1}) for the correlation parameter $\bar{C}_{ij}$, equation (\ref{eq: bethe Cij}) is  solved for the couplings $J_{ij}$. To calculate the local fields $h_i$, the solutions $J_{ij}$ are fed into the self-consistent equation (\ref{eq: bethe m}). Its solution in closed form completes the Bethe reconstruction.

The Bethe-Peierls approximation is related to a number of previous approaches. Expanding equation (\ref{eq: bethe Cij}) to first and second order in $J_{ij}$ recovers the naive mean-field and TAP reconstructions, respectively. Calculating the couplings and local fields using the self-consistent equations (\ref{eq: bethe m}) and (\ref{eq: bethe chi}) with $m_i=m_i^D$ and $\bar{C}_{ij}=C_{ij}^D$ estimated from data leads to the  independent-pair approximation~\cite{Roudi_FCN}.  On the other hand, inserting the estimated values $m_i^D$ and $C_{ij}^D$ into the places of $m_i$ and $C_{ij}$ as well as $\bar{C}_{ij}$ in equation (\ref{eq: bethe Cij})  (without using the solution  (\ref{eq: bethe chi 1}) for $\bar{C}_{ij}$),  one obtains the Sessak-Monasson reconstruction~\cite{Roudi_FCN,Sessak}. And lastly, since the extrema of Bethe-Peierls free energy are the fixed points of belief propagation \cite{Yedidia_15}, susceptibility propagation is expected to have a fixed point at the solution of Bethe reconstruction.

Interestingly, for a tree at zero field, the magnetizations $m_i$ are zero for finite system, and equations (\ref{eq: bethe Cij}) and (\ref{eq: bethe Cii}) take on a particularly simple form $(C^{-1})_{ij}= - \sh (J_{ij}) \ch (J _{ij})$ for $i \ne j$, and $(C^{-1})_{ii} = 1 + \sum_{j \in \partial i} \sh^2 ({J_{ij}})$ (see \cite{Eggarter} for the discussion on symmetry breaking in Cayley trees in the thermodynamic limit). This result can be understood as a geometrical relation, where the connected correlation has a simple interpretation: For any two nodes $i$ and $j$ in a tree there is a unique path $P_{ij}$ connecting them, and the correlation function can be written as $C_{ij} = \prod_{L_k \in P_{ij}} \th (J_{L_k})$ for $i \ne j$, and $C_{ii} = 1$, where $L_k$ denotes links in this path. Note that in the expression of $C_{ij}$, if one replaces the multiplication by summation over the links, the correlation matrix then becomes the distance matrix in the weighted tree with weight $W_{ij} = \th (J_{ij})$. A similar expression for the inverse of the distance matrix is already well-known in graph theory \cite{Bapat}. Bethe reconstruction then turns out to be particularly simple,
\begin{equation}
J_{ij} = - \frac{1}{2} \arcsh [2(C^{-1})_{ij}] \ , \ (i \ne j) \ .
\end{equation}

It is clear from the above description that the computational expense of Bethe reconstruction is mostly due to the inversion of the correlation matrix, as is the case for MF, TAP and SM. These methods are low computational expense methods, which are convenient to apply to systems of large sizes. The expense of inversion of the correlation matrix is of the order of $N^3$ for methods such as Gaussian elimination. On the other hand, the computation in a single update of SusP is already of the order of $N^3$, which makes SusP slower due to multiple iterative steps.

In the remainder of this note, we discuss the performance of Bethe reconstruction compared to MF, TAP, SM and SusP, which we refer to as the mean-field family. Although IP is faster than those methods, it usually leads to estimates of couplings inferior to all of these methods, so no comparison with IP is made.

We will first compare the performance of these methods for a tree and for the  Sherrington-Kirkpatrick (SK)-model~\cite{SK} in the absence of sampling noise (effectively infinite number of patterns). Then the effect of sampling noise is considered for a particular case of sparse random graphs. The comparisons proceed as follows: We first construct a graph of $N$ nodes, on which spins are located. Random values of couplings $J^0_{ij}$ are assigned to each edge, and random local fields $h^0_{i}$  are assigned at each spin. Two distributions for the random variables, the normal distribution with zero mean and standard deviation $\sigma$ ($\mathcal{N}(0,\sigma^2)$) and the uniform distribution on the interval $[a,b]$ ($\mathcal{U}(a,b)$), will be used (see the corresponding figure captions). The magnetizations $m_i^D$ and connected correlations $C_{ij}^D$ are calculated either by enumerating all the spin configurations (infinite sampling) or by performing Monte-Carlo simulation of the model at inverse temperature $\beta$ to generate a finite number of $M$ samples (finite sampling). The observed magnetizations and the connected correlations then serve as inputs to estimate the model parameters $\beta h_i$ and $\beta J_{ij} $ (where we explicitly reintroduced the inverse temperature). Concentrating on the reconstruction of the couplings, we measure the deviation of each solution from the underling couplings by the relative deviation 
\begin{equation}
d (J_{ij},J^0_{ij})= \left[\frac{{\sum_{i<j} (J_{ij} - J^0_{ij})^2}}{{\sum_{i<j} (J^0_{ij}) ^ 2}}\right]^{1/2}\ .
\end{equation}

\begin{center}
\begin{figure}[htp]
\begin{center}$
\begin{array}{cc}
(a) & (b)
\\
\includegraphics[width = 0.5 \textwidth]{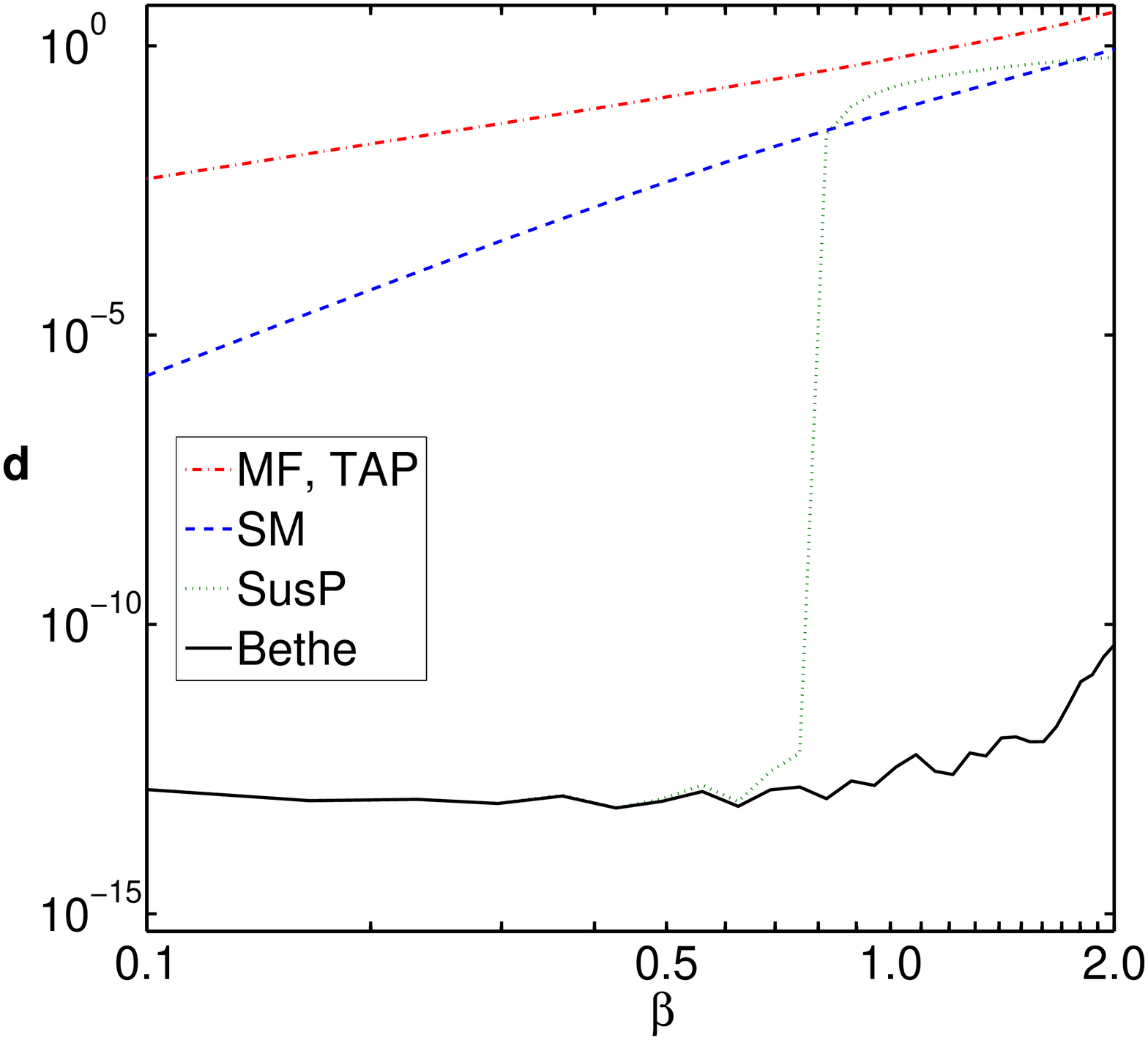}
&
\includegraphics[width = 0.5 \textwidth]{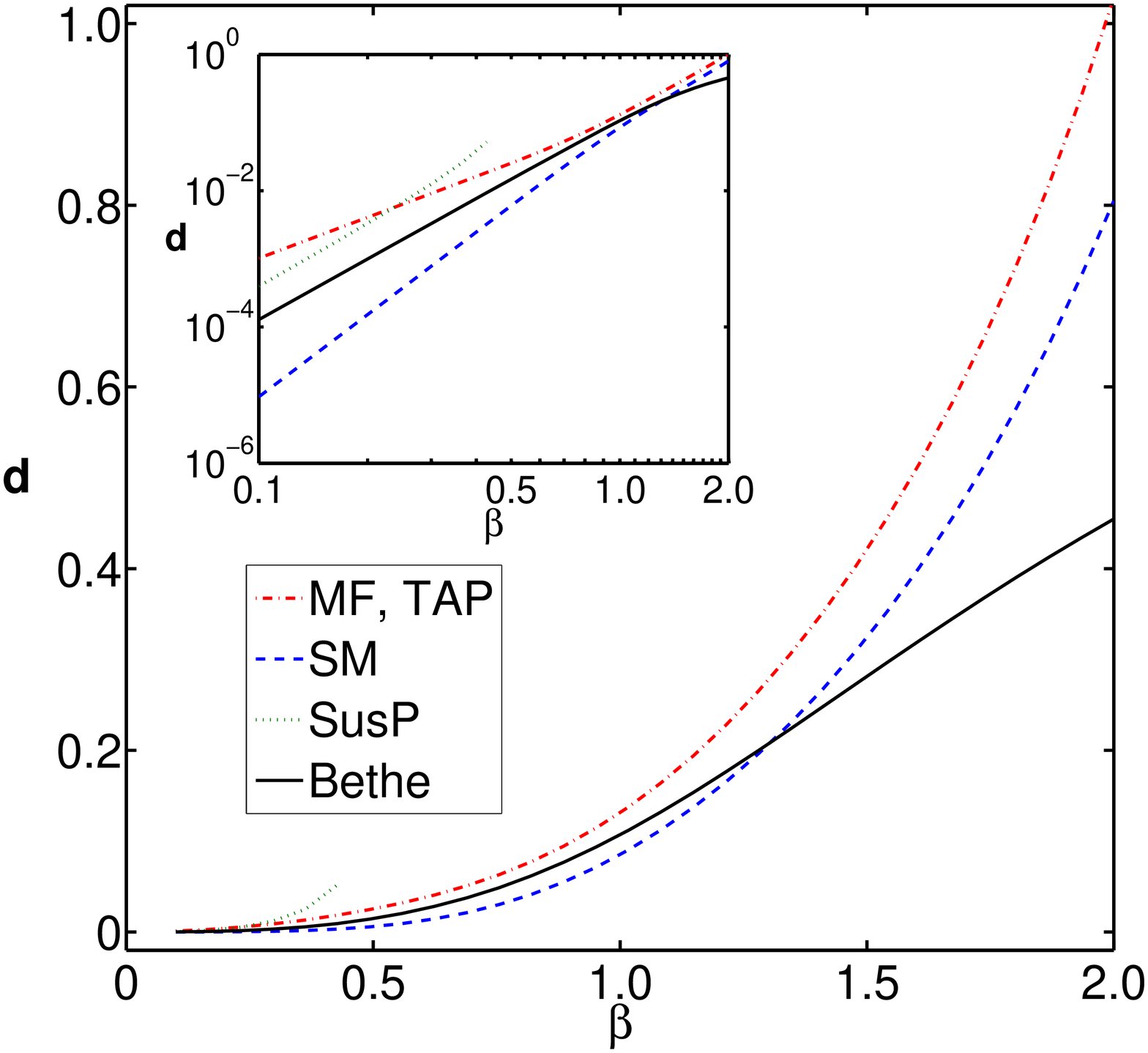}
\end{array}$
\end{center}
\caption{\textbf{Benchmarking in the infinite sampling limit.} Performance of naive mean-field (MF), TAP (TAP) (the same as MF in these cases), Sessak-Monasson (SM), susceptibility propagation (SusP) and Bethe reconstruction (Bethe), measured by the relative deviation $d$ (see main text) between the reconstructed couplings and the underlying couplings. (a) Cayley tree of connectivity $z=3$ over $N=22$ spins, $J^0_{ij} \sim \mathcal{U}(-1,+1)$, $h^0_i = 0$. (b) SK-model of $N=20$ spins, $J^0_{ij} \sim \mathcal{N}(0,1/N)$, $h^{0}_{i} = 0$. \label{fig: 1}}
\end{figure}
\end{center}

Figure \ref{fig: 1} (a) shows the reconstructions of the Ising model on a Cayley tree of connectivity $z=3$ with $N = 22$ spins. Here, the quality of Bethe reconstruction is limited only by the precision of numerical computations, confirming its exactness on a tree if the magnetizations $m^D_i$ and the connected correlations $C^{D}_{ij}$ are known exactly. SusP is also exact in this case, but is numerically less efficient and known to fail at low temperatures~\cite{Mezard_JOP}. On the other hand MF, TAP, and SM are all approximate, as seen particularly when the couplings are strong.

In figure \ref{fig: 1} (b) we turn to loopy graphs, and consider a particularly extreme case, the fully connected SK-model of $N = 20$ spins. At weak couplings (small $\beta$), the difference between the reconstruction of all the methods and the underling couplings are small, and in practice the differences between the methods will be obscured by sampling noise (see below). On the other hand, at strong couplings (larger $\beta$), Bethe reconstruction clearly  outperforms all the other methods of the mean-field family. Interestingly, even inside the glassy regime, a finite overlap between the Bethe-reconstructed couplings and the underlying couplings persists. Note that SusP also follows the line of Bethe reconstruction closely at weak couplings, but at strong couplings it fails to converge to the Bethe solution.

\begin{center}
\begin{figure}[htp]
\begin{center}$
\begin{array}{cc}
(a) & (b)
\\
\includegraphics[width = 0.5 \textwidth]{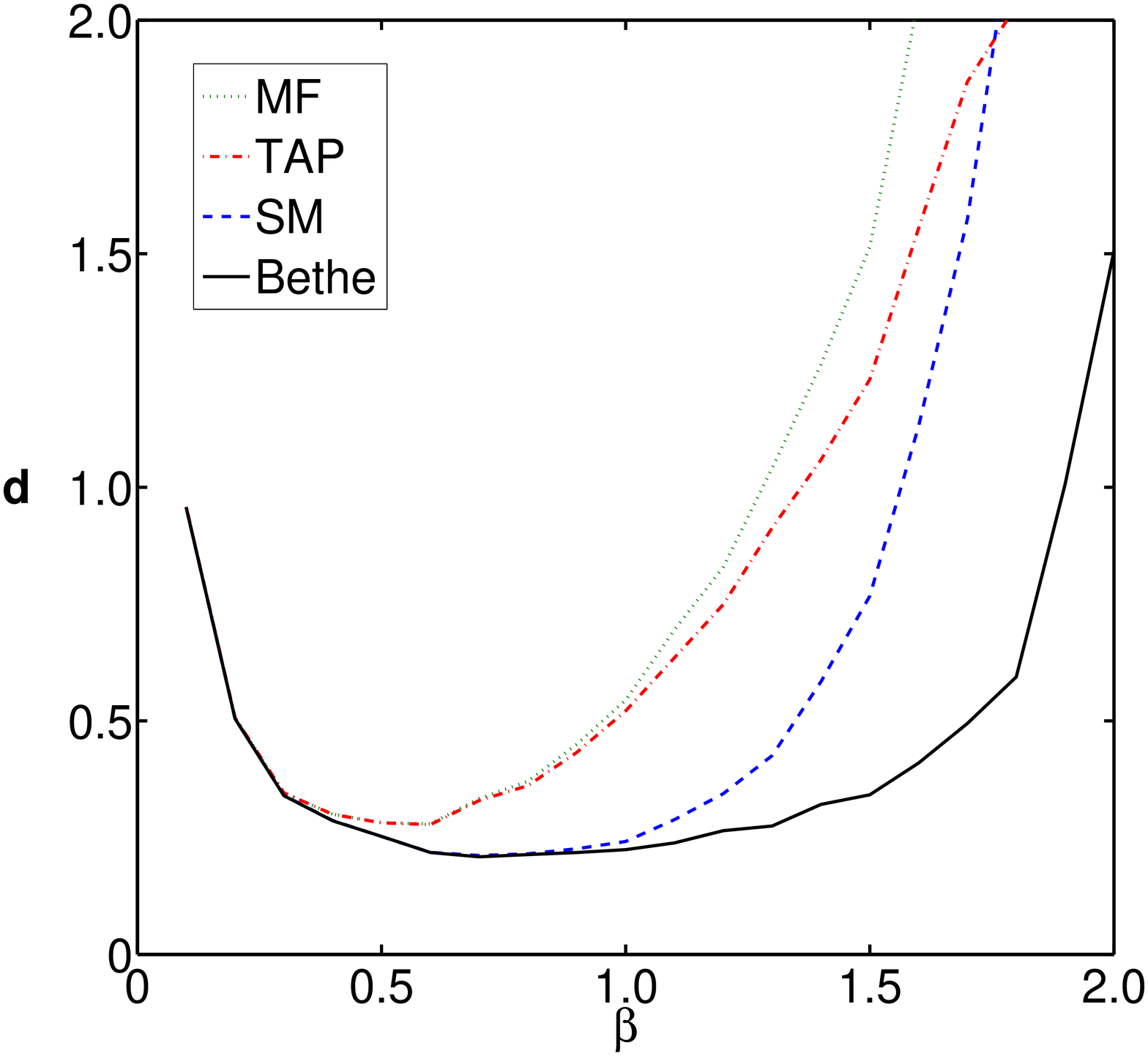}
&
\includegraphics[width = 0.5 \textwidth]{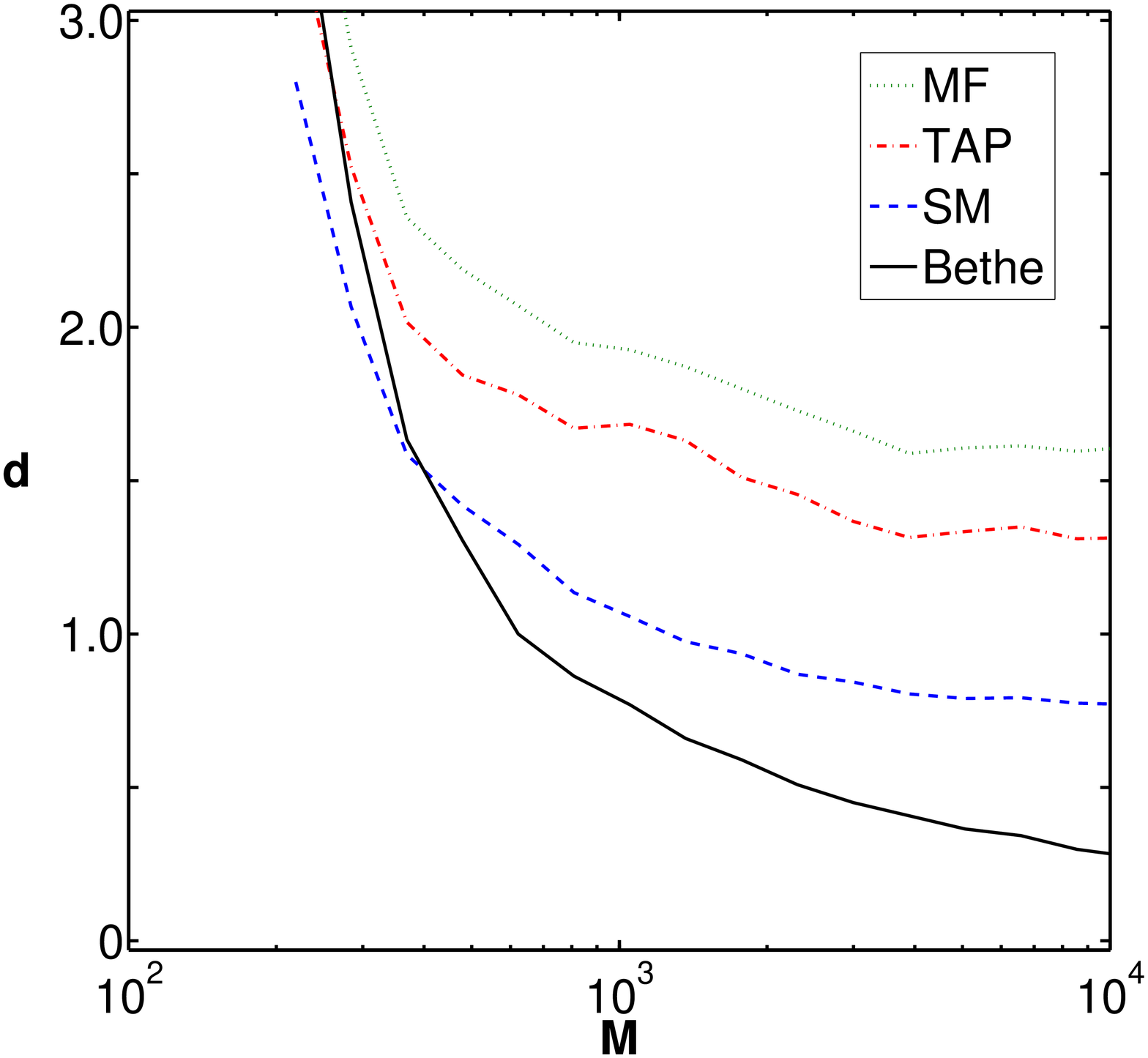}
\end{array}$
\end{center}
\caption{\textbf{Benchmarking in the case of finite sampling.} Performance of naive mean-field (MF), TAP (TAP), Sessak-Monasson (SM) and Bethe reconstruction (Bethe) on a random graph with fixed connectivity $z=3$, $N=50$ spins, $J^0_{ij} \sim \mathcal{U}(-1,+1)$, $h^0_{i} \sim \mathcal{U}(-0.1,+0.1)$. (a) The deviation $d$ versus the inverse temperature $\beta$, with $M = 5000$ samples. (b) The deviation $d$ versus the number of samples $M$ at $\beta = 1.5$. \label{fig: 2}}
\end{figure}
\end{center}

We now consider the Ising model with $N = 50$ spins placed on the vertices of a random graph with fixed connectivity $z = 3$~\cite{Viger}. Such graphs typically contain loops of length $\ln(N)$~\cite{Bollobas,Laumann}, but retain a locally tree-like structure. Figure \ref{fig: 2} (a) shows the results for a finite number of $M=5000$ samples, where we omit SusP from the comparison due to its numerical instability. At weak couplings (corresponding to high temperatures), reconstruction by any method is limited by sampling noise. Remarkably, the sampling noise tends to smear out the small difference between SM and Bethe reconstructions for the entire weak coupling regime. Again, the deviation $d$ of Bethe reconstruction increases only slowly through the strong coupling region, making it the best candidate for large couplings. Figure \ref{fig: 2} (b) probes the performances of the different methods at $\beta=1.5$ with varying number of samples. The results show that all the reconstruction methods from the mean-field family are sensitive to the sampling noise. For the present system, it requires more than $700$ samples to obtain a good reconstruction and to see a clear difference between the different methods.


In conclusion, we showed that the Bethe-Peierls approximation and linear response can serve as the basis for a method of the mean-field family to solve the inverse Ising problem. This Bethe reconstruction is not only computationally efficient but also stable through a wide range of couplings. While MF, TAP can be considered as low-order expansions in the couplings of Bethe reconstruction, the Sessak-Monasson reconstruction can be recovered by treating the parametric correlations in a particular manner. Although our observations suggest that susceptibility propagation and Bethe reconstruction are closely related, further work is needed to probe their relationship, in particular in the regime of strong couplings.

Since the Bethe-Peierls approximation works well for locally tree-like graphs, the extension of Bethe reconstruction to short-loop graphs, possibly made by linear response applied to the Kikuchi approximation \cite{Kikuchi,Yedidia_22,Tanaka}, is an interesting direction for future research. Another direction is extending the method to non-binary degrees of freedom, for which the linear response relation has been already studied~\cite{Welling_NC}.


\section*{Acknowledgment}
We would like to thank M. M\"{u}ller for useful discussions. We are grateful to M. Weigt, M. Welling, D. Saad and M. Opper for their suggestions. Comments of F. Klironomos, S. Ghozzi, A. Nourmohammad on the manuscript are acknowledged.  After submission of this manuscript we learnt of a manuscript by F. Ricci-Tersenghi containing results analogous to ours~\cite{Ricci}. This work was supported by Deutsche Forschungsgemeinschaft (DFG) grant SFB 680.


\newpage
\section*{References}
\bibliographystyle{unsrt}
\bibliography{references}
\end{document}